\documentclass[11pt,a4paper]{article}

\usepackage[utf8]{inputenc}
\usepackage[T1]{fontenc}
\usepackage{lmodern}
\usepackage[margin=2.5cm]{geometry}
\usepackage{amsmath,amssymb,amsfonts}
\usepackage{graphicx}
\usepackage{float}
\usepackage{enumitem}
\usepackage[numbers,sort&compress]{natbib}
\usepackage{authblk}
\usepackage[colorlinks=true,linkcolor=blue,citecolor=blue,urlcolor=blue]{hyperref}
\usepackage{orcidlink}
\usepackage[font=small,labelfont=bf,labelsep=period]{caption}

\title{Cooperation and frustration in the co-propagation of conflicting rumors on networks}

\author[1,2]{F.~Di~Ciocco\,\orcidlink{0009-0000-1528-9562}\thanks{Corresponding author: \href{mailto:fdiciocco@df.uba.ar}{fdiciocco@df.uba.ar}}}
\author[1,2]{L.~L.~Garcia\,\orcidlink{0009-0000-3313-0512}}
\author[3,4]{S.~C.~Ferreira\,\orcidlink{0000-0001-7159-2769}}
\author[1,2]{P.~Balenzuela\,\orcidlink{0000-0002-8581-4892}}

\affil[1]{Departamento de F\'isica, Facultad de Ciencias Exactas y Naturales, Universidad de Buenos Aires, 1428 Buenos Aires, Argentina}
\affil[2]{Instituto de F\'isica Interdisciplinaria y Aplicada (INFINA), CONICET -- Universidad de Buenos Aires, Buenos Aires, Argentina}
\affil[3]{Departamento de F\'isica, Universidade Federal de Vi\c{c}osa, 36570-900 Vi\c{c}osa, MG, Brazil}
\affil[4]{Instituto de Ci\^{e}ncias Matem\'{a}ticas e de Computa\c{c}\~{a}o, Universidade de S\~{a}o Paulo, 13566-590 S\~{a}o Carlos, SP, Brazil}

\date{}

\begin{document}

\maketitle

\begin{abstract}
Social media platforms expose users to volumes of information that often exceed their cognitive processing capacity. In contexts where forming an informed opinion is costly, individuals frequently rely on social reinforcement, validating information only if it is sufficiently confirmed by their local environment. To explore these dynamics, we propose a stochastic agent-based model with critical mass threshold to study the propagation of unverified information (rumors) in complex networks. In particular, we address the scenario in which agents are exposed to two correlated but opposed rumors about a specific topic. When the rumors are correlated, they exhibit cooperative propagation, meaning that exposure to one narrative facilitates the spread of the other. However, because the narratives are fundamentally opposed, we implement a microscopic rule whereby agents who adopt both rumors eventually discard them, raising the critical mass threshold for adopting future information. We interpret this mechanism as a representation of cognitive dissonance, capturing how psychological tension might be resolved through heightened skepticism. Through extensive numerical simulations, we identify three macroscopic regimes: single-rumor dominance; frustration, in which most agents become so skeptical that they no longer adopt either rumor, halting propagation across the network; and dynamic coexistence, in which a strong correlation keeps both rumors actively circulating throughout the network. Specifically, single-rumor dominance requires sufficiently fast dissonance resolution to break the symmetry enabled by correlation and shrinks as the critical mass threshold increment increases, whereas dynamic coexistence is governed solely by rumor correlation and dissonance resolution. To analytically characterize these behaviors, we develop a macroscopic dynamical description using a homogeneous mean-field approach and analyze the stability of the fixed points associated with the simulated regimes. In particular, we obtain dynamic coexistence analytically and show that it is stable whenever the correlation exceeds a threshold set by the dissonance rate, independently of the threshold increment, while numerical integration of the dynamical equations qualitatively reproduces the simulated regimes. Overall, these findings indicate that exposure to mutually reinforcing yet contradictory narratives in social media can drive communities toward narrative hegemony, widespread skepticism, or persistent circulation of conflicting claims.
\end{abstract}


\section{Introduction}\label{sec1}

The abundance of digital platforms and the diversification of communicative actors \cite{reuters2025digital} have significantly expanded the available supply of information. These environments, filled with overwhelming amounts of data, interact with cognitive mechanisms that shape how users select, process, and evaluate the information they receive, generating diverse patterns of news consumption \cite{cicchini2022news, flamino2023political, robertson2023negativity, galeazzi2024unveiling}. In these saturated environments, individuals frequently encounter competing narratives whose veracity is costly or impossible to verify independently.

When exposed to such conflicting narratives, the resulting psychological tension can be conceptualized through the lens of cognitive dissonance theory \cite{festinger1957dissonance}. This framework posits that the discomfort of holding incompatible beliefs motivates strategies to restore internal coherence \cite{harmon_jones_mills1999dissonance}. Furthermore, because the absolute veracity of these narratives cannot be independently assessed, individuals heavily rely on their local social environment to validate and adopt information in the first place. This reliance is driven by two main channels: normative influence, where individuals adjust their opinions to gain social approval or group belonging, and informational influence, where they assume their peers possess superior knowledge about the ambiguous issue \cite{deutsch1955study}. This necessity for social reinforcement when adopting unverified information is a hallmark of complex contagion \cite{guilbeault2018complex}, in which multiple peer confirmations are required to establish a belief when the information is contested or lacks objective verification.

These dynamics of social validation have been modeled extensively. In a seminal work, Granovetter proposed a threshold model \cite{granovetter1978threshold} applicable to diverse collective behaviors, including rumor diffusion. In this framework, an individual adopts a belief only once the fraction of peers who have already done so exceeds a personal critical mass threshold, giving rise to nonlinear collective dynamics and cascade effects. Building on this premise, subsequent works have extended threshold mechanics to analyze processes of conformity and consensus formation in complex networks, highlighting how local interaction structures and individual heterogeneity govern the spread of unverified information \cite{centola2007complex, centola2010spread, Watts_Model}.

A distinct line of research models rumor diffusion by analogy to infectious diseases, adapting classical epidemiological frameworks \cite{maia2023controversy, takiguchi2026siscompetitionmodelconflicting}. Within this framework, the propagation of multiple competing rumors has been studied under the assumption that agents either remain neutral or adopt a single rumor exclusively \cite{he2012influence, trpevski2010model, kostka2008word}, while more recent approaches incorporate temporal dynamics, weighting rumors influence by their persistence across repeated exposures \cite{zarezade2017correlated, farajtabar2016multistage}. Among them, agent-based models on complex networks have proven particularly suited to capturing how network structure and local interactions give rise to collective patterns of diffusion \cite{bessi2015science, delvicario2016spreading, zanette2002dynamics}. In all of these models, however, agents are restricted to a single belief state at any given time.

Based on the threshold model introduced by Watts \cite{Watts_Model} in which each node on a network adopts a state once the fraction of its neighbors in that state meets or exceeds an individual threshold, potentially triggering global cascades from a small seed, we propose a rumor-spreading framework in which agents are simultaneously exposed to two opposing rumors concerning a single subject. When the veracity of information cannot be independently verified, agents adopt a rumor only if a sufficient fraction of their neighbors subscribe to it. The interplay between these narratives is governed by a correlation parameter: the higher the correlation, the more the attention drawn by one narrative amplifies the circulation of the other. Crucially, our model allows an agent to spread both rumors simultaneously, a state of cognitive dissonance that prevents the formation of a stable opinion. To resolve this tension, the agent follows a specific microscopic rule: it eventually discards both rumors and increases its critical mass threshold, requiring a stronger social reinforcement to adopt either rumor in the future. We interpret this mechanism through the lens of cognitive dissonance, as a way to restore internal coherence through heightened skepticism.

This capacity for agents to adopt both conflicting rumors simultaneously represents the primary innovation of our framework. We argue that allowing concomitant rumors' spread is relevant for the study of information diffusion, as individuals frequently retain and transmit multiple narratives concurrently. Furthermore, enabling simultaneous adoption is necessary to analyze the dynamics of cooperative, correlated rumors: restricting agents to a single belief state imposes an artificial exclusion rule that forces a competitive dynamic, even among complementary rumors.

The primary objective of this study is to determine the conditions under which a single rumor achieves system-wide propagation, ultimately dominating public opinion. Through extensive numerical simulations, we identify three macroscopic regimes governed by the correlation between the rumors and the rate at which agents resolve their dissonance: single-rumor dominance, in which one narrative saturates the network; frustration, in which repeated rejections raise critical mass thresholds until most agents become too skeptical to adopt either rumor, halting propagation; and dynamic coexistence, in which a strong correlation keeps both rumors actively circulating throughout the network. Furthermore, we show that the magnitude of the critical mass threshold increase upon belief rejection determines the extent of the dominance region in the parameter space, while the dynamic coexistence regime is insensitive to it. Finally, we derive a set of macroscopic dynamical equations via a homogeneous mean-field approximation, analyze the stability of their fixed points, obtaining an analytical expression for the dynamic coexistence state and the minimum correlation required to sustain it, and show that their numerical integration qualitatively reproduces the regimes observed in the agent-based model. Beyond the theoretical characterization of the model, these results illustrate how social reinforcement coupled with dissonance-induced skepticism can push online information ecosystems either toward consensus around a single unverified claim, toward collective disengagement and distrust, or toward the sustained co-circulation of contradictory narratives.

\section{The Model}\label{sec2}

We propose a modification of the linear threshold model on a complex network, initially inspired by the work of Duncan Watts \cite{Watts_Model} on adoption cascades. The original Watts model describes the dynamics of idea or technology adoption, positing that each agent \textit{i} possesses a skepticism or resistance to adoption, represented by a critical mass threshold $\varphi_i$ \cite{granovetter1978threshold}. This threshold defines the minimum \textit{fraction} of the agent's neighbors who must have adopted the idea for the agent to also adopt it. In the original Watts model, agents with $\varphi_i = 0$ represent early adopters who adopt the innovation without requiring any external influence, while those with $\varphi_i = 1$ represent stubborn agents who only adopt the innovation if all of their neighbors have already adopted it.

\begin{figure}[H]
    \centering
    \includegraphics[width=\textwidth]{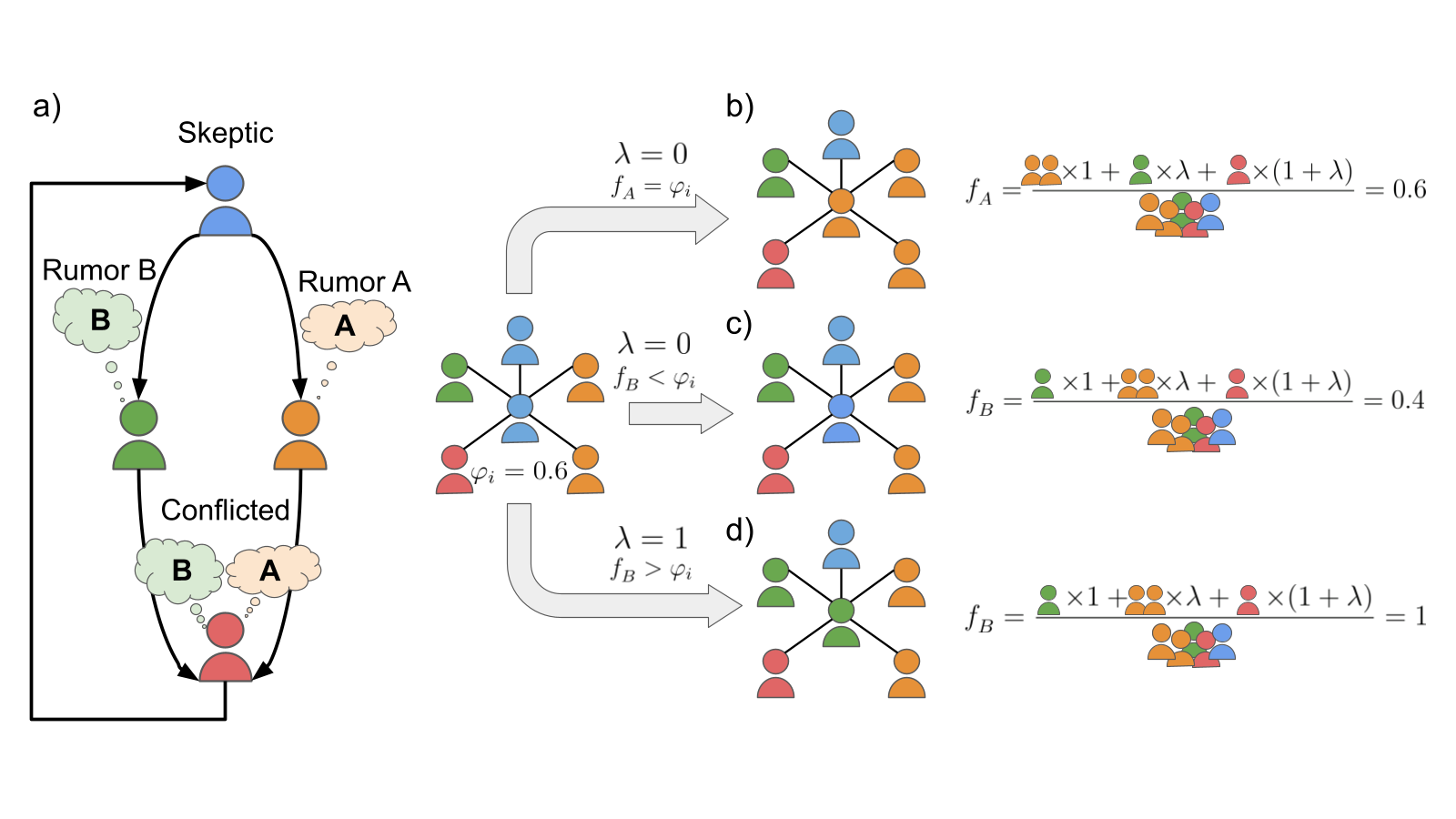}
    \caption{\textbf{Description of the model states and dynamics.} (a) Agents can be in one of four states: ``Skeptic'', ``Rumor A'', ``Rumor B'' and ``Conflicted''. These states are defined by the rumors the agents have adopted. Agents can only transition between states along the paths indicated by the arrows. For example, an agent in the ``Skeptic'' state can only transition to either the ``Rumor A'' or the ``Rumor B'' state, depending on the rumor adopted. From these single-rumor states, it transitions to the ``Conflicted'' state upon adopting the second rumor. Finally, from this state, it discards both rumors and returns to the ``Skeptic'' state. (b-d) An agent adopts a rumor if its cumulative influence $f$ of believing neighbors meets or exceeds its threshold value. For a threshold value of $\varphi_i = 0.6$, we show the scenarios where an agent attempts to adopt rumors A and B with no correlation between them, succeeding in the former but failing in the latter. We also illustrate the case where the agent attempts to adopt rumor B with full correlation between rumors, this time successfully adopting the new narrative}
    \label{fig:Model_description}
\end{figure}

Our model adapts this framework to study rumor propagation when objective verification is unavailable. Agents are exposed to rumors about a subject and, lacking full information or the capacity to process it, they decide to believe a rumor only if a sufficient fraction of their neighbors already believe it, i.e., it requires a critical mass threshold \cite{granovetter1978threshold}. We model the spread of two distinct rumors within the network, which we denote as a positive rumor (describing benefits or virtues) and a negative rumor (describing defects or detriments)---for example, claims that a given habit is relaxing versus claims that it damages one's health. Agents can therefore be found in one of four possible states: \textbf{Skeptic} (adopted neither rumor), \textbf{Rumor A} (adopted only rumor A), \textbf{Rumor B} (adopted only rumor B), or \textbf{Conflicted} (adopted both rumors). Fig.~\ref{fig:Model_description}(a) shows these states and the cycle that an agent follows: from believing in no rumor, to believing in one of them, to believing in both, and back to believing in none.

The interaction between the two rumors is governed by a correlation parameter $\lambda \in [0,1]$. The boundary cases are: $\lambda = 0$, which means no correlation between the rumors, and $\lambda = 1$, which means total correlation. When rumors are correlated, the propagation of one of the rumors promotes the propagation of the other. An agent \textit{i} with $k$ neighbors attempting to adopt rumor \textit{r} receives, besides the influence of all neighbors also adopting $r$ with weight $1/k$, an influence from neighbors who believe in the \textit{correlated} rumor \textit{r}', weighted by a factor $\lambda/k$. For an agent \textit{i} with threshold $\varphi_i$ trying to adopt rumor \textit{r}, the adoption is successful if the \textit{cumulative influence} $f_r$ of neighbors believing in \textit{r} meets or exceeds $\varphi_i$. Notably, because this additive social pressure accounts for overlapping beliefs, the cumulative influence can exceed $1$, a value impossible in a single-rumor model, but which reflects the synergistic reinforcement in our framework. As in the Watts model, critical mass thresholds are restricted to $[0,1]$, as they represent a fraction of neighbors.

For example, consider an agent \textit{i} with five neighbors and a threshold $\varphi_i=0.6$. In the case of no correlation ($\lambda = 0$), as shown in Fig.~\ref{fig:Model_description}(b), to adopt rumor A, the agent only counts neighbors explicitly believing in A (those in the ``Rumor A'' or ``Conflicted'' states). There are three of its five neighbors that believe in A, so the cumulative influence of neighbors that adopted rumor A is $f_A = 0.6$. Since $0.6 \ge \varphi_i$, the agent adopts A. Conversely, if we consider the agent trying to adopt rumor B as in Fig.~\ref{fig:Model_description}(c), the agent only has two neighbors that believe in rumor B and thus the cumulative influence is $f_B = 0.4$. Therefore the agent does not adopt rumor B, as $0.4 < \varphi_i$. However, this result changes under total correlation ($\lambda = 1$), as shown in Fig.~\ref{fig:Model_description}(d). When the agent attempts to adopt rumor B, neighbors who believe in rumor A now also contribute. Neighbors in the ``Rumor B'' state count once, neighbors in the ``Rumor A'' state count with weight $\lambda$, and neighbors in the ``Conflicted'' state count $1+\lambda$ times, once for each rumor. With one neighbor in ``Rumor B'', two in ``Rumor A'' and one in ``Conflicted'', the cumulative influence is $f_B = (1 + 2\lambda + (1+\lambda))/5 = 1$. Since $1 > \varphi_i$, the agent adopts B.

Belief in both positive and negative rumors generates cognitive dissonance---a psychological tension arising from conflicting views about the subject \cite{festinger1957dissonance}. To restore cognitive coherence, the agent facing this conflict (in the ``Conflicted'' state) eventually abandons both rumors, returning to the ``Skeptic'' state. This transition is governed by the parameter $\alpha$, the probability that an agent in the ``Conflicted'' state abandons both rumors, so the higher $\alpha$, the faster agents leave the ``Conflicted'' state. Furthermore, to model an agent that becomes more strict in their criteria for readoption, their skepticism is increased by raising their critical mass threshold $\varphi_i$ by an amount $\delta$ after dropping the rumors. If an agent's threshold would increase above 1, it is then set to 1. That is, the threshold is updated as $\varphi_i \leftarrow \min(\varphi_i+\delta, 1)$. Note that an agent with $\varphi_i = 1$ is not excluded from the dynamics: it can still adopt a rumor whenever its cumulative influence reaches or exceeds 1.

\section{Model simulation}\label{sec3}

We simulate the propagation of rumors on an Erd\H{o}s-R\'enyi network \cite{erdos1959random} with $N=100,000$ agents and mean degree $\langle k \rangle = 8$. We explore the parameter space defined by $\lambda \in [0,1]$ and $\alpha \in (0,1]$, with a fixed threshold increment $\delta = 0.01$, and average the results over 100 independent realizations for each parameter combination. At each time step, an agent and a rumor $r$ are selected at random, and the agent is updated according to the following rules:

\begin{itemize}
    \item \textbf{Adoption:} If the agent does not already spread rumor $r$, it adopts it if the cumulative influence of its neighbors believing in $r$ meets or exceeds its critical mass threshold ($f_r \ge \varphi_i$).
    \item \textbf{Dissonance:} If the agent already spreads rumor $r$ and also spreads the opposing rumor (i.e., it is in the ``Conflicted'' state), it abandons both rumors with probability $\alpha$, transitioning to the ``Skeptic'' state. Upon this transition, its critical mass threshold increases by $\delta$; if the increment would raise the threshold above 1, the threshold is set to 1. If instead the agent only spreads the chosen rumor $r$, nothing happens and the simulation advances to the next step.
\end{itemize}

One unit of time corresponds to $N$ such updates. The simulation proceeds until the system reaches an equilibrium, either because no further transitions occur (frozen state, in which the rumors no longer propagate) or because the fraction of agents in each state no longer changes (stationary state). The initial critical mass threshold of each agent is drawn from a normal distribution with mean $\mu=0.3$ and standard deviation $\sigma = 0.1$, truncated to the interval $[0,1]$. The system is initialized by randomly selecting $10\%$ of the agents to believe in rumor A and, independently, $10\%$ to believe in rumor B. Since these sets may overlap, a small fraction of agents may believe in both rumors initially.

\begin{figure}[H]
    \centering
    \includegraphics[width=0.9\textwidth]{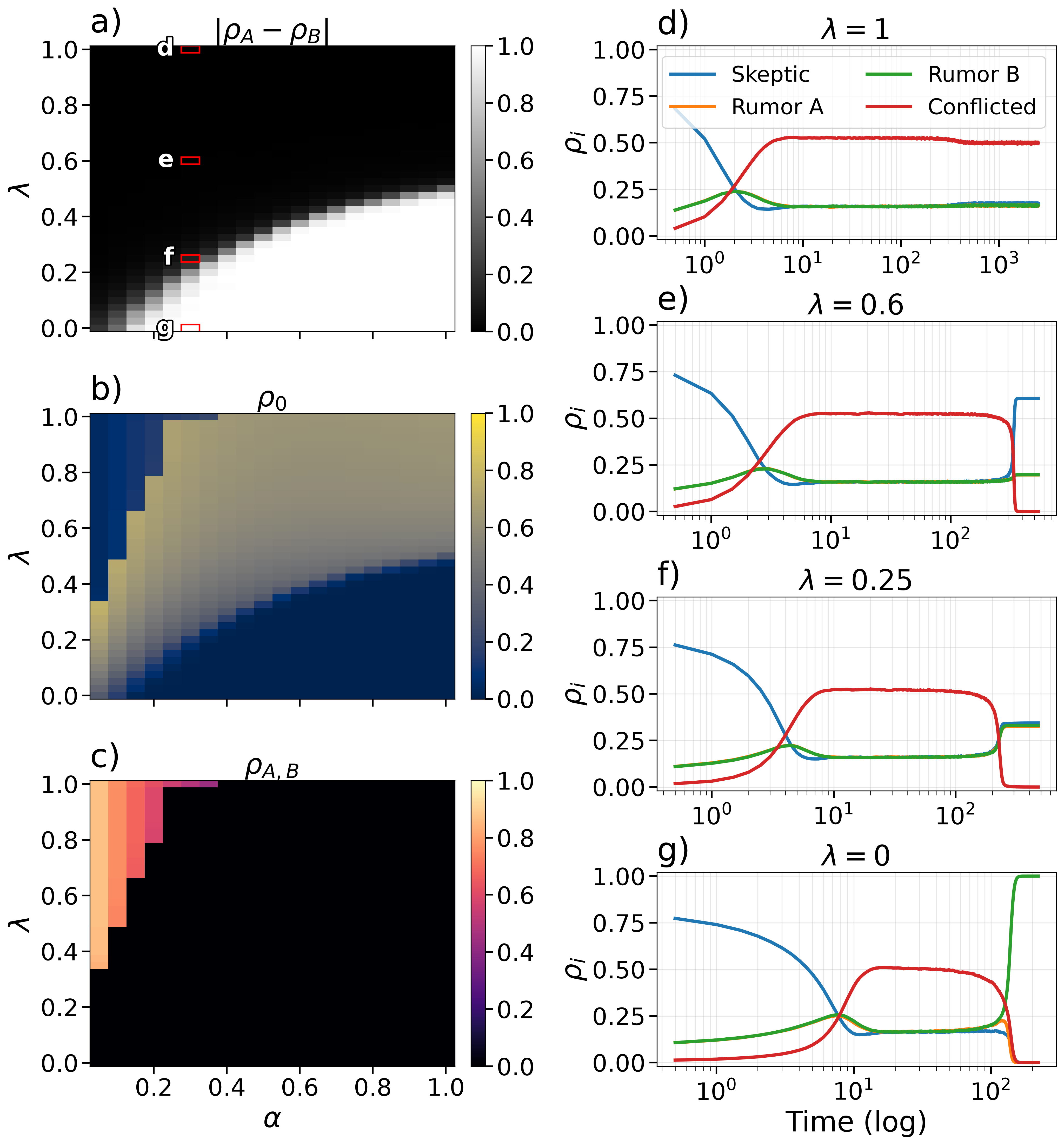}
    \caption{\textbf{Simulation results.} We performed 100 simulations for each point in the $\lambda-\alpha$ parameter space using an Erd\H{o}s-R\'enyi network ($N=100,000$, $\langle k \rangle = 8$) with fixed $\delta = 0.01$. Panels on the left column show mean values at the end of the simulations for: (a) the absolute difference between rumor fractions, $|\rho_A - \rho_B|$, revealing a region of single-rumor dominance for low correlation and high dissonance; (b) the fraction of agents in the ``Skeptic'' state, $\rho_0$, indicating a frustration region, between the dominance and the dynamic coexistence regions, in which the majority of agents fall into the ``Skeptic'' state; (c) the fraction of agents in the ``Conflicted'' state, $\rho_{A,B}$, which remains non-zero only for low $\alpha$ and high $\lambda$ (dynamic coexistence region). Panels on the right column show the time evolution, in logarithmic scale, of the fraction of agents in each state for a single realization with fixed $\alpha = 0.3$, at the points marked with red squares in (a): (d) dynamic coexistence ($\lambda = 1$); (e) frustration ($\lambda = 0.6$); (f) transition boundary ($\lambda = 0.25$); (g) single-rumor dominance ($\lambda = 0$)}
    \label{fig:Simulation_results}
\end{figure}

In our model, each agent who adopts a rumor increases the exposure of its neighbors, so that the small initial seed can set off a chain of adoptions that spreads through the network, i.e., a global cascade. When both rumors trigger global cascades simultaneously, their overlapping diffusion drives a large part of the population into the ``Conflicted'' state. Since agents leave this state with probability $\alpha$, they eventually revert to the ``Skeptic'' state, raising their critical mass thresholds in the process. This reset mechanism induces a cyclic propagation dynamics with three possible macroscopic outcomes. First, the symmetry between the rumors may break, allowing a single rumor to dominate the network, a regime we call single-rumor dominance. Second, the successive threshold increments may halt propagation altogether, leaving the ``Skeptic'' state as the most populated one and only smaller fractions adhering to either rumor, a regime we call frustration. Finally, the cycle may never end, keeping a large fraction of the population in the ``Conflicted'' state indefinitely, a regime we call dynamic coexistence. We study how the parameters $\lambda$ and $\alpha$ determine which of these outcomes the system reaches.

In Fig.~\ref{fig:Simulation_results}(a), we plot the ensemble average of the absolute difference between the final rumor fractions, $|\rho_A-\rho_B|$, across the $\lambda-\alpha$ parameter space. A value of $1$ indicates the complete dominance of a single rumor, whereas $0$ indicates that both rumors end up with the same fraction of agents. This panel therefore separates the single-rumor dominance regime from the regimes in which neither rumor prevails, which are distinguished in panels (b) and (c). For low correlation $\lambda$, dominance emerges across a broad range of dissonance values $\alpha$. As the correlation increases, however, the dissonance probability required to achieve single-rumor dominance grows non-linearly. This transition is governed by the lifespan of the ``Conflicted'' state: a high dissonance $\alpha$ shortens the time that agents spend in this state, severely reducing its capacity to spread the rumors. Consequently, any initial statistical asymmetry between the rumors is rapidly amplified, triggering a cascade in which one rumor saturates the network and the other is driven to extinction through the transient ``Conflicted'' state. Conversely, when the dissonance is low, the ``Conflicted'' state accumulates a substantial population, and a high correlation prevents either rumor from gaining an advantage, since agents believing in one rumor also contribute to the adoption of the other. The system is then locked into a symmetric cycle of activation and deactivation, which ends in either the frustration or the dynamic coexistence regime.

To further characterize the final state of the system, Fig.~\ref{fig:Simulation_results}(b) shows the ensemble average of the final ``Skeptic'' fraction, $\rho_0$. This fraction is negligible in the single-rumor dominance regime, where one rumor saturates the network, and small in the dynamic coexistence regime, discussed below. Between these two regions lies the frustration regime, where the ``Skeptic'' state becomes the most populated one, with a fraction that increases with correlation and decreases with dissonance, controlled by $\lambda$ and $\alpha$, respectively. The remaining agents are trapped in either the ``Rumor A'' or the ``Rumor B'' state. They adopted a single rumor, but their local neighborhood lacked sufficient influence either to propagate that belief further or to induce the adoption of the opposing rumor, so they never entered the ``Conflicted'' state. Thus, skeptical agents and single-rumor adopters coexist statically in a frozen configuration: rather than sustaining active co-propagation, both rumors survive only within isolated clusters of believers while most of the population has abandoned their spread.

Fig.~\ref{fig:Simulation_results}(c) shows the ensemble average of the final fraction of agents in the ``Conflicted'' state, $\rho_{A,B}$. This fraction vanishes in both the dominance and the frustration regimes, but it is large for low dissonance and high correlation. In this dynamic coexistence regime, the system never reaches an absorbing state: the fractions of agents in each state become stationary, but agents keep cycling between them. This happens because the repeated threshold increments bring them to the maximum value of 1 for practically all agents. As discussed in Section~\ref{sec2}, such agents can still adopt a rumor when their cumulative influence reaches or exceeds 1, which a high correlation makes possible, since neighbors believing in either rumor contribute to it. A low dissonance, in turn, keeps a large population in the ``Conflicted'' state, whose agents count towards both rumors, sustaining the cumulative influences needed to readopt them.

Figures~\ref{fig:Simulation_results}(d--g) illustrate the temporal evolution of the fractions of agents in each state for fixed $\alpha=0.3$ across the four regimes, indicated as squares in Figure~\ref{fig:Simulation_results}(a): dynamic coexistence ($\lambda=1$), frustration ($\lambda=0.6$), the transition boundary ($\lambda=0.25$), and single-rumor dominance ($\lambda=0$). In all cases, the initial cascades of both rumors quickly drive about half of the population into the ``Conflicted'' state. The height of this plateau is set by the dissonance parameter $\alpha$, decreasing as $\alpha$ increases, and barely depends on $\lambda$. The regimes differ in how this plateau ends. In the dynamic coexistence regime, it never ends: the fractions remain stationary, with $\rho_{A,B}\approx 0.5$ and the rest of the population evenly split among the other three states. In the frustration regime, the plateau collapses after a few hundred time units, once the rising thresholds halt propagation, and most agents end in the ``Skeptic'' state while the rest split symmetrically between the ``Rumor A'' and ``Rumor B'' states. In the dominance regime, the plateau is shorter and ends asymmetrically: one rumor grows at the expense of the other and saturates the network as the ``Conflicted'' fraction vanishes. At the transition boundary, where the outcome fluctuates across realizations, the ``Skeptic'' fraction competes directly with the single-rumor populations; in the realization shown in Fig.~\ref{fig:Simulation_results}(f), all three states end with comparable fractions ($\approx 1/3$), marking the crossover toward frustration as increasing $\lambda$ progressively favors final skepticism.

We now analyze the effect of the threshold increment $\delta$, which sets how strongly agents penalize conflicting information. When $\delta \approx 0$, agents barely adjust their critical mass thresholds after each deactivation and can readopt the rumors many times. Conversely, a high $\delta$ characterizes a ``strict'' community, in which a few episodes of dissonance suffice to make agents highly resistant to adoption. Since thresholds cannot exceed 1, $\delta$ essentially determines how many deactivations an agent undergoes before its threshold saturates, and therefore how fast propagation is suppressed.

\begin{figure}[H]
    \centering
    \includegraphics[width=\textwidth]{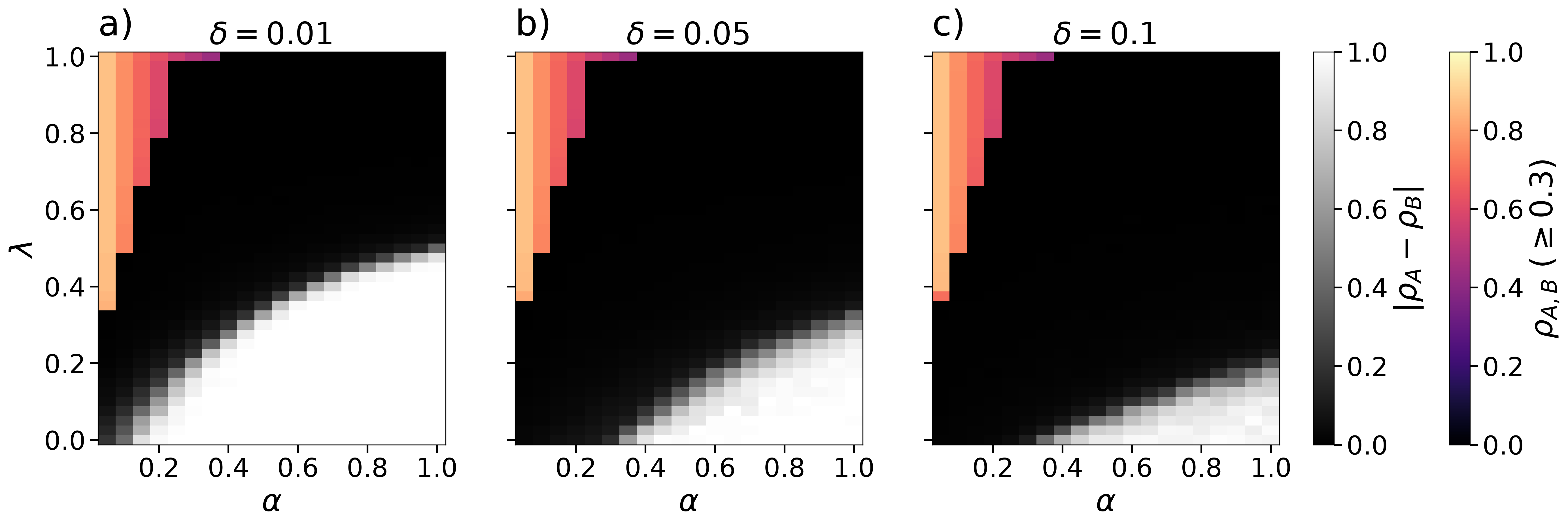}
    \caption{\textbf{Effect of $\delta$ on model behavior.} Final state of the model in the $\lambda-\alpha$ parameter space for (a) $\delta=0.01$, (b) $\delta=0.05$ and (c) $\delta=0.1$, averaged over 100 simulations for each point using an Erd\H{o}s-R\'enyi network ($N=100,000$, $\langle k \rangle = 8$). The grayscale shows the mean absolute difference between rumor fractions, $|\rho_A - \rho_B|$: light areas correspond to the single-rumor dominance region and dark areas to the frustration region. Overlaid in color is the mean fraction of agents in the ``Conflicted'' state, $\rho_{A,B}$, shown only where $\rho_{A,B} \geq 0.3$, which identifies the dynamic coexistence region (upper left). As the threshold penalty $\delta$ increases, the dominance region shrinks towards high $\alpha$ and low $\lambda$, while the dynamic coexistence region remains almost unchanged}
    \label{fig:Delta_effect}
\end{figure}

Fig.~\ref{fig:Delta_effect} shows the final state of the model in the $\lambda-\alpha$ parameter space for $\delta=0.01$, $0.05$ and $0.1$. As $\delta$ increases, the single-rumor dominance region shrinks, surviving only for high dissonance and low correlation, and the frustration regime takes over the black part of the parameter space. The dynamic coexistence region, shown in color scale in the top left part, is in contrast practically insensitive to $\delta$.

The contraction of the dominance region reflects the time that symmetry breaking requires. As discussed above, dominance arises from the amplification of initial asymmetries between the rumors. When $\delta$ is low, agents that revert to the ``Skeptic'' state retain a high probability of readopting a rumor, giving fluctuations enough time to grow until one rumor out-competes the other. When $\delta$ is high, the rapid escalation of critical mass thresholds halts propagation before the symmetry breaks, and the system ends in the frustration regime. Only for high dissonance and low correlation, where the ``Conflicted'' state is short-lived and asymmetries are amplified quickly, does dominance persist for $\delta=0.1$. In the area lost by the dominance regime, propagation stops at an earlier stage than in the rest of the frustration region: the ``Skeptic'' fraction is lower
and both rumors retain larger groups of believers.

The dynamic coexistence regime, on the other hand, is insensitive to $\delta$ because the thresholds of practically all agents saturate at 1 for any value of $\delta$, which only changes how quickly saturation is reached. Once the thresholds are saturated, the stationary state no longer depends on $\delta$: the colored region in the three panels of Fig.~\ref{fig:Delta_effect} is practically identical, and the final fractions of agents in each state are practically the same for the three values of $\delta$. Similarly, in the frustration regime at high correlation, a larger $\delta$ halts propagation earlier but barely modifies the final ``Skeptic'' fraction.

Overall, single-rumor dominance relies on a delicate balance between the synergistic spreading enabled by rumor correlation and the suppression of propagation imposed by dissonance and the threshold penalty, whereas dynamic coexistence is determined by $\lambda$ and $\alpha$ alone.

\section{Mean-field description}\label{sec4}

\subsection{Dynamical equations}\label{subsec1}

We describe the dynamical behavior of the model by proposing a set of ordinary differential equations for the fractions of agents in each of the four states under a homogeneous mean-field analysis. We define the state of the system using four variables representing the fractions of agents in each state: the fraction of ``Skeptic'' agents ($\rho_0$), the fraction adopting only rumor A ($\rho_A$), the fraction adopting only rumor B ($\rho_B$), and the fraction of ``Conflicted'' agents ($\rho_{A,B}$). For this analysis, we assume that every agent observes, on average, the same environment; thus, the fraction of neighbors in each state $s$ is equal to the global fraction of agents in that state, $\rho_s$.

The rates at which populations change are given by the mass-action law, where the total population is assumed to be constant. We begin by constructing the mean-field equation for $\rho_0$. The ``Skeptic'' state loses agents when they adopt either rumor A or rumor B, and gains agents when those in the ``Conflicted'' state drop both rumors due to cognitive dissonance. Therefore, to construct the equation for $\dot \rho_0$, we consider three contributing terms:

\begin{enumerate}
    \item \textbf{Loss due to adoption of rumor A:} This term represents the rate at which ``Skeptic'' agents transition to the ``Rumor A'' state. The probability of a ``Skeptic'' agent encountering the rumor is proportional to the cumulative influence of neighbors transmitting it, $f_A$ (Section~\ref{sec2}), which in the mean-field approximation reads
    \begin{equation}
        f_A = \rho_A + \lambda \, \rho_B + (1+\lambda) \, \rho_{A,B}. \label{f_A_mf}
    \end{equation}
    Here, $\rho_A$ contributes the direct influence from agents who believe only in rumor A, $\rho_B$ contributes the cross-influence from agents who believe only in rumor B, weighted by the correlation $\lambda$, and $\rho_{A,B}$ contributes the influence from agents believing in both rumors, who count once as believers in A and $\lambda$ times as believers in B. Since $f_A$ can exceed 1 when $\lambda > 0$ whereas critical mass thresholds lie in $[0,1]$, we cap the cumulative influence at 1, replacing $f_A$ by $\min(f_A, 1)$ wherever it appears in the dynamical equations.

    However, an encounter results in adoption only if the social pressure meets or exceeds the agent's threshold. We define $\omega_A(t)$ as the fraction of agents whose critical mass threshold is lower than the current cumulative influence $f_A$, which acts as the probability that an agent can adopt rumor A. Since thresholds increase when agents drop both rumors, $\omega_A$ depends on time. Combining the encounter rate with the adoption probability, the loss term for this transition is $\omega_A \, \rho_0 \, f_A$.

    \item \textbf{Loss due to adoption of rumor B:} Symmetrically, the rate at which ``Skeptic'' agents adopt rumor B depends on the cumulative influence of neighbors transmitting B, which reads
    \begin{equation}
        f_B = \rho_B + \lambda \, \rho_A + (1+\lambda) \, \rho_{A,B}. \label{f_B_mf}
    \end{equation}
    Defining $\omega_B(t)$ analogously, the loss term for rumor B is $\omega_B \, \rho_0 \, f_B$.

    \item \textbf{Gain from cognitive dissonance:} The ``Skeptic'' state gains agents when those in the ``Conflicted'' state abandon both rumors to resolve their psychological tension. This transition occurs at a fixed rate $\alpha$, yielding a gain term $\alpha \, \rho_{A,B}$.
\end{enumerate}

Summing these three terms, the differential equation for $\rho_0$ reads
\begin{equation*}
    \dot \rho_0 = - \omega_A \, \rho_0 \, f_A - \omega_B \, \rho_0 \, f_B + \alpha \, \rho_{A,B}.
\end{equation*}

The equations for the other three fractions are constructed using similar reasoning. The fraction of agents believing only in rumor A, $\rho_A$, increases when ``Skeptic'' agents adopt rumor A and decreases when agents in the ``Rumor A'' state adopt rumor B, thus becoming ``Conflicted''; the equation for $\rho_B$ follows symmetrically by interchanging A and B. Finally, the fraction of agents believing in both rumors, $\rho_{A,B}$, increases when agents holding a single belief adopt the complementary one, and decreases when they drop both rumors due to cognitive dissonance. The dynamics of the four fractions in the mean-field approximation is therefore given by

\begin{align}
    \frac{d\rho_0}{dt} &= - \omega_A \, \rho_0 \, f_A - \omega_B \, \rho_0 \, f_B + \alpha \, \rho_{A,B}, \label{rho_0} \\
    \frac{d\rho_A}{dt} & = \omega_A \, \rho_0 \, f_A - \omega_B \, \rho_A \, f_B, \label{rho_A} \\
    \frac{d\rho_B}{dt} & = \omega_B \, \rho_0 \, f_B - \omega_A \, \rho_B \, f_A, \label{rho_B} \\
    \frac{d\rho_{A,B}}{dt} & = \omega_B \, \rho_A \, f_B + \omega_A \, \rho_B \, f_A - \alpha \, \rho_{A,B}, \label{rho_{A,B}}
\end{align}

where $f_A$ and $f_B$ are given by Eqs.~\eqref{f_A_mf} and \eqref{f_B_mf}. For readability, we do not write the cap explicitly; it must be kept in mind that each cumulative influence $f_A$ and $f_B$ (as well as each bracket below) is replaced by 1 whenever it exceeds this value.

Since every agent is in one of the four states, the sum of all fractions is conserved, that is, $\rho_0 + \rho_A + \rho_B + \rho_{A,B} = 1$, as can easily be verified by summing Eqs.~\eqref{rho_0}--\eqref{rho_{A,B}}.

This constraint allows us to reduce the system from four coupled equations to three. Using the condition $\rho_{A,B} = 1 - \rho_0 - \rho_A - \rho_B$ to eliminate $\rho_{A,B}$ from Eqs.~\eqref{rho_0}--\eqref{rho_B} yields the reduced system

\begin{align}
    \frac{d\rho_0}{dt} &= - \omega_A \, \rho_0 \, [(1+\lambda) \, (1-\rho_0) - \lambda \, \rho_A - \rho_B] - \omega_B \, \rho_0 \, [(1+\lambda) \, (1-\rho_0) - \lambda \, \rho_B - \rho_A] \nonumber \\
    &\qquad + \alpha \, (1 - \rho_0 - \rho_A - \rho_B), \label{rho_0_vf} \\
    \frac{d\rho_A}{dt} & = \omega_A \, \rho_0 \, [(1+\lambda) \, (1-\rho_0) - \lambda \, \rho_A - \rho_B] - \omega_B \, \rho_A \, [(1+\lambda) \, (1-\rho_0) - \lambda \, \rho_B - \rho_A], \label{rho_A_vf} \\
    \frac{d\rho_B}{dt} & = \omega_B \, \rho_0 \, [(1+\lambda) \, (1-\rho_0) - \lambda \, \rho_B - \rho_A] - \omega_A \, \rho_B \, [(1+\lambda) \, (1-\rho_0) - \lambda \, \rho_A - \rho_B]. \label{rho_B_vf}
\end{align}

\subsection{Fixed-point stability analysis}\label{subsec2}

The complexity of the dynamical equations \eqref{rho_0_vf}--\eqref{rho_B_vf} makes a full analytical determination of all fixed points challenging. We therefore focus on identifying several physically relevant equilibrium states, associated with the regimes observed in the agent-based simulations of Section~\ref{sec3}, and on analyzing their stability.

We define the state vector as $\vec{\rho} = (\rho_0, \rho_A, \rho_B)$, where $\rho_{A,B}$ is implicitly determined by the normalization condition. The equilibrium states analyzed are:

\begin{enumerate}
    \item \textbf{Complete skepticism}: All agents eventually return to the ``Skeptic'' state, having abandoned both rumors. The corresponding vector is $\vec{\rho}^{\,0} = (1,0,0)$.
    \item \textbf{Single-rumor dominance}: One rumor spreads through the entire network, while the other dies out. The corresponding vectors are $\vec{\rho}^{\,A} = (0,1,0)$ and $\vec{\rho}^{\,B} = (0,0,1)$.
    \item \textbf{Frustration}: Critical mass thresholds become so high that the rumors can no longer spread. Since agents leave the ``Conflicted'' state at rate $\alpha > 0$, this state eventually becomes extinct, so that $\rho_{A,B} = 0$. The corresponding vector is $\vec{\rho}^{\,F} = (\rho_0^F, \rho_A^F, \rho_B^F)$, with $\rho_0^F + \rho_A^F + \rho_B^F = 1$, which corresponds to a surface of fixed points in the three-dimensional phase portrait.
    \item \textbf{Dynamic coexistence}: The thresholds of practically all agents reach their maximum value of 1, but agents keep cycling between states, so that a finite fraction of the population remains in the ``Conflicted'' state. The corresponding vector is $\vec{\rho}^{\,C} = (\rho_0^C, \rho_A^C, \rho_B^C)$, with $\rho_{A,B}^C = 1 - \rho_0^C - \rho_A^C - \rho_B^C > 0$.
\end{enumerate}

First, we analyze the case of \textbf{complete skepticism}. Substituting $\vec{\rho}^{\,0} = (1,0,0)$ into Eqs.~\eqref{rho_0_vf}--\eqref{rho_B_vf} immediately yields $\dot{\rho}_0 = \dot{\rho}_A = \dot{\rho}_B = 0$. Thus, $\vec{\rho}^{\,0}$ is an equilibrium point regardless of the values of $\omega_A$, $\omega_B$ and $\alpha$.

Physical intuition suggests that this state is unstable: if the equilibrium is perturbed by a single agent adopting a rumor, the system will either remain where it is (if the agent cannot influence its neighbors) or cascade away from the equilibrium, but it will not return to it. To prove this instability formally, we analyze the Jacobian matrix of the system. We simplify the analysis by considering the case in which both rumors evolve symmetrically, that is, $\rho_A = \rho_B \equiv \rho$ and $\omega_A = \omega_B \equiv \omega$. The system then reduces to the two coupled equations

\begin{align}
    \frac{d\rho_0}{dt} &= - 2 \omega \, \rho_0 \, (1+\lambda) \, (1-\rho_0- \rho) + \alpha \, (1 - \rho_0 - 2 \rho), \label{rho_0_reducido} \\
    \frac{d\rho}{dt} & = \omega \, (1+\lambda) \, (\rho_0 - \rho) \, (1-\rho_0 - \rho). \label{rho_reducido}
\end{align}

The Jacobian matrix $J$ of this reduced system is

\begin{align*}
    J = \begin{pmatrix} \frac{\partial \dot\rho_0}{\partial\rho_0} & \frac{\partial \dot\rho_0}{\partial\rho} \\ \frac{\partial \dot\rho}{\partial\rho_0} & \frac{\partial \dot\rho}{\partial\rho} \end{pmatrix} = \begin{pmatrix} -2 \omega (1+\lambda)(1-2\rho_0-\rho)-\alpha & 2 \omega (1+\lambda)\rho_0 - 2\alpha \\ \omega (1+\lambda)(1-2\rho_0) & \omega (1+\lambda)(2\rho-1) \end{pmatrix}.
\end{align*}

The eigenvalues of $J$ evaluated at $\vec{\rho}^{\,0}$, i.e., at $(\rho_0, \rho) = (1,0)$, are $\gamma_1 = \omega (1+\lambda)$ and $\gamma_2 = -\alpha$. Since all parameters are positive, we have that $\gamma_1$ is non-negative while $\gamma_2$ is negative. Therefore, $\vec{\rho}^{\,0}$ is a saddle point and an unstable equilibrium.

Next, we examine the \textbf{single-rumor dominance} state, taking $\vec{\rho}^{\,A} = (0,1,0)$ as an example. We can easily verify from Eqs.~\eqref{rho_0_vf}--\eqref{rho_B_vf} that for $\vec{\rho}^{\,A}$ to be a fixed point, we require either $\lambda = 0$ or $\omega_B = 0$. Under the condition $\omega_B = 0$, meaning that no agent can be induced to adopt rumor B, the ``Rumor B'' state eventually becomes extinct: as rumor A spreads through the network, agents in the ``Rumor B'' state adopt it and become ``Conflicted'', then revert to the ``Skeptic'' state and finally adopt rumor A, since they can no longer be induced to adopt rumor B. Keeping all three equations in Eqs.~\eqref{rho_0_vf}--\eqref{rho_B_vf} with $\omega_B = 0$ and evaluating the three-dimensional Jacobian matrix $J$ at $\vec{\rho}^{\,A}$, i.e., at $(\rho_0, \rho_A, \rho_B) = (0,1,0)$ where $f_A = 1$, yields the reducible matrix

\begin{align*}
    \left. J \right|_{\vec{\rho}^{\,A}} = \begin{pmatrix} \frac{\partial\dot\rho_0}{\partial\rho_0} & \frac{\partial\dot\rho_0}{\partial\rho_A} & \frac{\partial\dot\rho_0}{\partial\rho_B} \\ \frac{\partial\dot\rho_A}{\partial\rho_0} & \frac{\partial\dot\rho_A}{\partial\rho_A} & \frac{\partial\dot\rho_A}{\partial\rho_B} \\ \frac{\partial\dot\rho_B}{\partial\rho_0} & \frac{\partial\dot\rho_B}{\partial\rho_A} & \frac{\partial\dot\rho_B}{\partial\rho_B} \end{pmatrix}_{\vec{\rho}^{\,A}} = \begin{pmatrix} -\omega_A - \alpha & \;\;-\alpha\;\; & -\alpha \\ \omega_A & \;\;0\;\; & 0 \\ 0 & \;\;0\;\; & -\omega_A \end{pmatrix},
\end{align*}

with eigenvalues $\gamma_1 = -\alpha$, $\gamma_2 = -\omega_A$, and $\gamma_3 = -\omega_A$. So, the fixed point $\vec{\rho}^{\,A}$ is stable under the conditions $\omega_B = 0$ and $\omega_A > 0$. Analogously, $\vec{\rho}^{\,B}$ corresponds to a stable fixed point under the conditions $\omega_A = 0$ and $\omega_B > 0$.

In the case of \textbf{frustration}, where $\omega_A = \omega_B = 0$, Eqs.~\eqref{rho_0_vf}--\eqref{rho_B_vf} imply that $\rho_A$ and $\rho_B$ are constant while $\dot{\rho}_0 = \alpha \, (1 - \rho_0 - \rho_A - \rho_B)$, so the equilibrium condition $\dot{\rho}_0 = 0$ is satisfied by any vector $\vec{\rho}^{\,F} = (\rho_0^F, \rho_A^F, \rho_B^F)$ with $\rho_0^F + \rho_A^F + \rho_B^F = 1$, which is stable since $\partial \dot \rho_0 / \partial \rho_0 = -\alpha < 0$. This indicates that, once thresholds become prohibitive, the system converges to $\vec{\rho}^{\,F}$ as the remaining agents in the ``Conflicted'' state revert to the ``Skeptic'' state.

Finally, we consider the case of \textbf{dynamic coexistence}. As shown in Section~\ref{sec3}, in this regime the repeated threshold increments bring the thresholds of practically all agents to their maximum value of 1, yet agents keep adopting both rumors because the cumulative influences reach their cap of 1. We therefore assume that $f_A$ and $f_B$ are capped at 1, so that $\omega_A = \omega_B = 1$ and every adoption term reduces to the fraction of agents that can adopt the rumor (e.g., $\omega_A \, \rho_0 \, f_A = \rho_0$). Substituting these conditions into Eqs.~\eqref{rho_0_vf}--\eqref{rho_B_vf} yields

\begin{align}
    \frac{d\rho_0}{dt} &= -2\rho_0 + \alpha \, (1 - \rho_0 - \rho_A - \rho_B), \label{rho_0_coex} \\
    \frac{d\rho_A}{dt} & = \rho_0 - \rho_A, \label{rho_A_coex} \\
    \frac{d\rho_B}{dt} & = \rho_0 - \rho_B. \label{rho_B_coex}
\end{align}

The fixed point $\vec{\rho}^{\,C}$ follows from setting these time derivatives to zero. The last two equations yield $\rho_A^C = \rho_B^C = \rho_0^C$, and substituting this result into the first one gives

\begin{align}
    \rho_0^C = \rho_A^C = \rho_B^C = \frac{\alpha}{2 + 3\alpha}, \qquad \rho_{A,B}^C = 1 - 3\rho_0^C = \frac{2}{2 + 3\alpha}. \label{rho_C}
\end{align}

This solution is consistent with our assumptions only if the cumulative influences actually reach the cap at $\vec{\rho}^{\,C}$, that is, if $f_A = f_B \ge 1$. Since $\rho_A^C = \rho_B^C = \rho_0^C$, the cumulative influence at this point is $f_A = (1+\lambda)(\rho_0^C + \rho_{A,B}^C)$, and the condition becomes

\begin{align}
    \lambda \ge \frac{2\alpha}{2+\alpha}. \label{lambda_coex}
\end{align}

Therefore, the fraction of agents in the ``Conflicted'' state at $\vec{\rho}^{\,C}$ depends only on the dissonance rate $\alpha$, whereas the correlation determines whether this state can be sustained: dynamic coexistence requires a correlation above a minimum value that increases with $\alpha$.

To determine the stability of $\vec{\rho}^{\,C}$, we compute the Jacobian matrix of Eqs.~\eqref{rho_0_coex}--\eqref{rho_B_coex},

\begin{align*}
    J = \begin{pmatrix} \frac{\partial\dot\rho_0}{\partial\rho_0} & \frac{\partial\dot\rho_0}{\partial\rho_A} & \frac{\partial\dot\rho_0}{\partial\rho_B} \\ \frac{\partial\dot\rho_A}{\partial\rho_0} & \frac{\partial\dot\rho_A}{\partial\rho_A} & \frac{\partial\dot\rho_A}{\partial\rho_B} \\ \frac{\partial\dot\rho_B}{\partial\rho_0} & \frac{\partial\dot\rho_B}{\partial\rho_A} & \frac{\partial\dot\rho_B}{\partial\rho_B} \end{pmatrix} = \begin{pmatrix} -2 - \alpha & \;\;-\alpha\;\; & -\alpha \\ 1 & \;\;-1\;\; & 0 \\ 1 & \;\;0\;\; & -1 \end{pmatrix}.
\end{align*}

Its eigenvalues are $\gamma_1 = -1$, associated with perturbations in which $\rho_A$ and $\rho_B$ change in opposite directions, and
\begin{align*}
    \gamma_{2,3} = \frac{-(3+\alpha) \pm \sqrt{\alpha^2 - 6\alpha + 1}}{2}.
\end{align*}
For $\alpha < 3 - 2\sqrt{2} \approx 0.17$, $\gamma_2$ and $\gamma_3$ are real and negative, whereas for larger values of $\alpha$ they are complex conjugates with negative real part $-(3+\alpha)/2$, so the system approaches $\vec{\rho}^{\,C}$ through damped oscillations. In both cases, all eigenvalues have negative real parts, and therefore $\vec{\rho}^{\,C}$ is a stable fixed point whenever it exists, that is, whenever Eq.~\eqref{lambda_coex} is satisfied.

\subsection{Dynamical integration}\label{subsec3}

To validate the proposed dynamical description and the stability analysis of the fixed points, we numerically integrated the mean-field equations \eqref{rho_0_vf}--\eqref{rho_B_vf} to study the system behavior as a function of the parameters $\alpha$ and $\lambda$, with $\delta$ fixed. Since the adoption probabilities $\omega_A$ and $\omega_B$ depend on time, they must be recalculated at every time step of the integration, which requires an explicit expression for them.

Following their definition in Section~\ref{subsec1}, we calculate $\omega_A$ and $\omega_B$ as the cumulative distribution function (CDF) of the time-dependent threshold distribution $\Phi(\varphi,t)$, evaluated at the current cumulative influences (capped at 1), that is,

\begin{align}
    \omega_A (t) & = \int_{0}^{f_A} \Phi(\varphi,t) \, d\varphi, \label{omega_a} \\
    \omega_B (t) & = \int_{0}^{f_B} \Phi(\varphi,t) \, d\varphi. \label{omega_b}
\end{align}

We model $\Phi(\varphi,t)$ as a normal distribution with a time-dependent parameter $\mu(t)$ and a fixed standard deviation $\sigma$, truncated to the window $[\mu(t)-\mu_0, 1]$, where $\mu_0 = \mu(0)$. Initially, this window is $[0,1]$, so that $\Phi$ coincides with the threshold distribution of the agent-based model (Section~\ref{sec3}). The upper bound remains fixed at 1, since thresholds cannot exceed this value. The lower bound, in turn, moves with $\mu(t)$, as the lowest thresholds also increase when agents abandon the rumors. The truncation also has a dynamical role: since $\Phi$ vanishes below the lower bound, $\omega_r$ is exactly zero whenever $f_r$ falls below it, for any rumor $r$. As shown in Section~\ref{subsec2}, this condition is required for the existence of the stable fixed points in which the rumors stop spreading. Conversely, since $\Phi$ vanishes above 1, $\omega_r = 1$ whenever $f_r$ reaches its cap, as assumed for the dynamic coexistence fixed point.

As agents transition from the ``Conflicted'' state to the ``Skeptic'' state, they increment their thresholds by $\delta$. Since this deactivation process occurs stochastically across the population, we assume that it shifts the distribution without altering its shape. The evolution of $\mu(t)$ is therefore driven by the flux of agents leaving the ``Conflicted'' state ($\alpha \, \rho_{A,B}$), scaled by the threshold increment $\delta$, so that

\begin{align}
    \dot \mu = \delta \, \alpha \, \rho_{A,B}. \label{Mu_dynamic}
\end{align}

Note that $\mu(t)$ is the location parameter of the underlying normal distribution, not the mean of the truncated one. Once $\mu(t)$ exceeds 1, the distribution piles up against the upper bound, reproducing the saturation of thresholds at 1 observed in the agent-based model.

We integrated the system formed by Eqs.~\eqref{rho_0_vf}--\eqref{rho_B_vf} and \eqref{Mu_dynamic}, with $\omega_A$ and $\omega_B$ given by Eqs.~\eqref{omega_a}--\eqref{omega_b}, using a fourth-order Runge--Kutta method. We explored the parameter space $\lambda \in [0,1]$ and $\alpha \in [0.05,1]$, with $\delta = 0.01$ and the threshold distribution initialized with $\mu_0 = 0.3$ and $\sigma = 0.1$, as in the agent-based simulations. We verified that extending the integration time does not change the results. The initial condition mimics that of the agent-based model: each rumor is initially adopted, independently, by a fraction $p_A = 0.1 + \epsilon$ and $p_B = 0.1 - \epsilon$ of the agents, so that $\rho_A(0) = p_A (1-p_B)$, $\rho_B(0) = p_B (1-p_A)$, $\rho_{A,B}(0) = p_A p_B$ and $\rho_0(0) = (1-p_A)(1-p_B)$. The small asymmetry $\epsilon = 10^{-3}$ breaks the symmetry of the equations, allowing one rumor to dominate the other; as a consequence, whenever a single rumor dominates, it is rumor A.

\begin{figure}[H]
    \centering
    \includegraphics[width=0.9\textwidth]{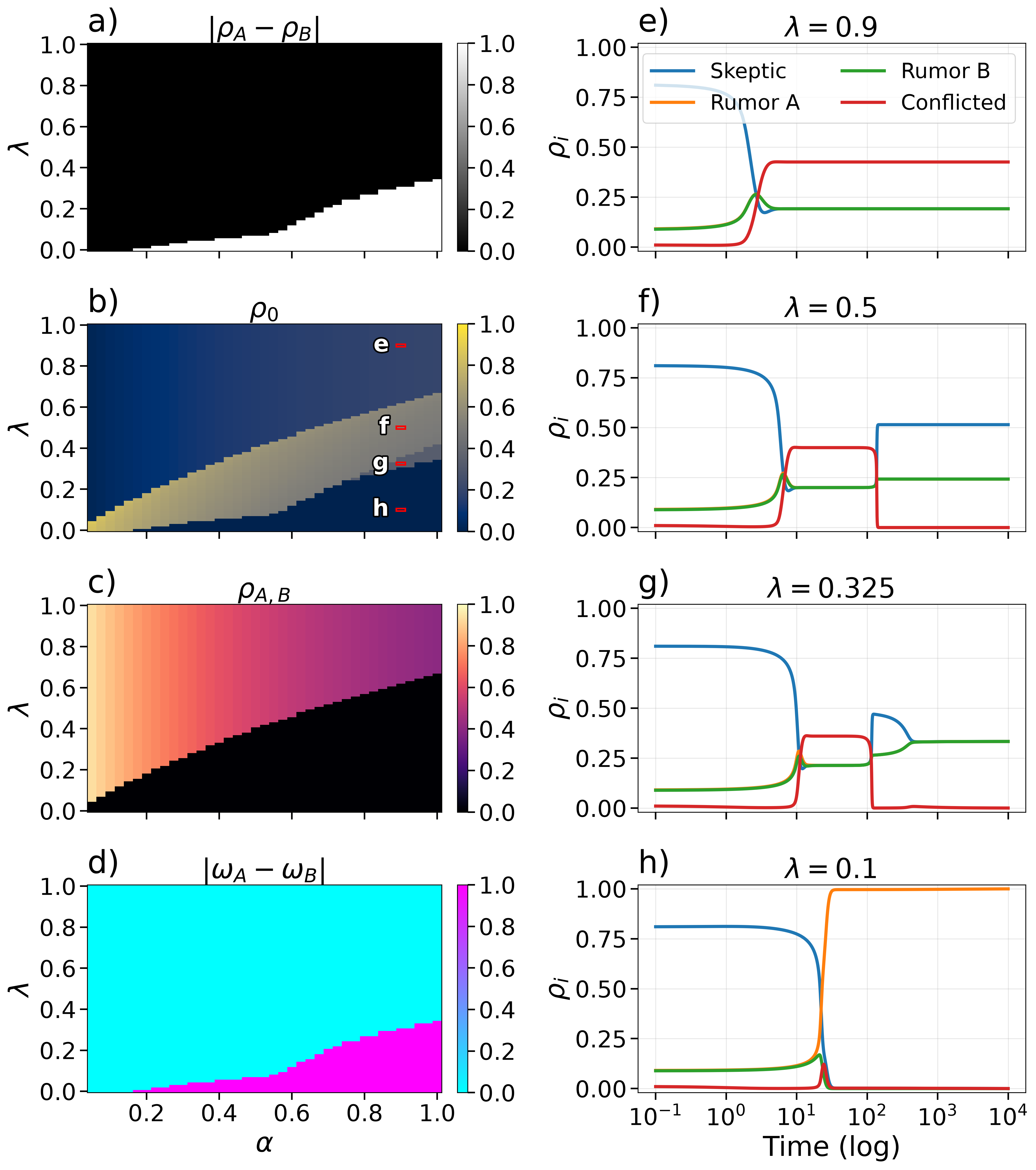}
    \caption{\textbf{Numerical integration of the homogeneous mean-field equations.} Panels on the left column show the final values in the $\lambda-\alpha$ parameter space, with fixed $\delta = 0.01$, of: (a) the absolute difference between rumor fractions, $|\rho_A - \rho_B|$, revealing a region of single-rumor dominance for low correlation and high dissonance; (b) the fraction of agents in the ``Skeptic'' state, $\rho_0$, which distinguishes the dynamic coexistence region (top), the frustration region (intermediate band) and, for high dissonance, a narrow region with $\rho_0 = 1/3$ between the frustration and dominance regions; (c) the fraction of agents in the ``Conflicted'' state, $\rho_{A,B}$, which is non-zero only in the dynamic coexistence region, where it depends only on $\alpha$; (d) the absolute difference between the adoption probabilities, $|\omega_A - \omega_B|$, which is close to 1 in the dominance region and 0 elsewhere. Panels on the right column show the time evolution, in logarithmic scale, of the fraction of agents in each state with fixed $\alpha = 0.9$, at the points marked with red squares in (b): (e) dynamic coexistence ($\lambda = 0.9$); (f) frustration ($\lambda = 0.5$); (g) slow convergence to the equipartitioned state $\rho_0 = \rho_A = \rho_B = 1/3$ ($\lambda = 0.325$); (h) single-rumor dominance ($\lambda = 0.1$)}
    \label{fig:Dynamic_integration}
\end{figure}

Fig.~\ref{fig:Dynamic_integration}(a) shows the absolute difference between the final rumor fractions, $|\rho_A - \rho_B|$. As in the agent-based model, a single-rumor dominance region appears for low correlation and high dissonance, while in the rest of the parameter space both rumors end up with the same fraction of agents. Fig.~\ref{fig:Dynamic_integration}(d) shows that the dominance region coincides with the region where $|\omega_A - \omega_B| \approx 1$, that is, where the dominant rumor can be adopted by every agent ($\omega_A = 1$) and the other cannot be adopted at all ($\omega_B = 0$). This agrees with the stability analysis of Section~\ref{subsec2}, which showed that $\vec{\rho}^{\,A}$ is a stable fixed point when $\omega_B = 0$ and $\omega_A > 0$. The time evolution in this regime, shown in Fig.~\ref{fig:Dynamic_integration}(h) for $\lambda = 0.1$, is fast: after the initial cascade, rumor A saturates the network while the ``Rumor B'' and ``Conflicted'' fractions vanish.

The fractions of agents in the ``Skeptic'' and ``Conflicted'' states, shown in Figs.~\ref{fig:Dynamic_integration}(b) and (c), separate the region in which neither rumor prevails into the frustration and dynamic coexistence regimes. In the dynamic coexistence regime, which occupies the upper part of the parameter space, a large fraction of the population remains in the ``Conflicted'' state, while the remaining agents split evenly among the other three states. These final fractions coincide with those of the fixed point $\vec{\rho}^{\,C}$ given by Eq.~\eqref{rho_C}, and the lower boundary of the region is given by Eq.~\eqref{lambda_coex}. Fig.~\ref{fig:Dynamic_integration}(e) shows that, for $\lambda = 0.9$, the system reaches this stationary state shortly after the initial cascades and remains there with $\omega_A = \omega_B = 1$, as assumed in the derivation of $\vec{\rho}^{\,C}$. Between this region and the dominance region lies the frustration regime, where the ``Conflicted'' state is emptied and the ``Skeptic'' state becomes the most populated one, with the rest of the population split symmetrically between the ``Rumor A'' and ``Rumor B'' states. As shown in Fig.~\ref{fig:Dynamic_integration}(f) for $\lambda = 0.5$, the ``Conflicted'' fraction first reaches a plateau, similar to that of the dynamic coexistence regime, which collapses once the rising thresholds make $\omega_A = \omega_B = 0$. The system then remains at a frustration fixed point $\vec{\rho}^{\,F}$.

Overall, the integration qualitatively reproduces the three regimes observed in the agent-based simulations, as well as their arrangement in the parameter space. The extent of the regions, however, differs; in particular, the dynamic coexistence region extends to high dissonance values, where the simulations show frustration, since according to Eq.~\eqref{lambda_coex} the mean-field description only requires $\lambda \ge 2/3$ to sustain dynamic coexistence at $\alpha = 1$. The differences are indeed expected, since the mean-field equations become exact only in the limit of a complete graph, where every individual is connected to every other individual. For finite connectivity, dynamical correlations between pairs of connected individuals start to play an important role \cite{mata2014heterogeneous}. More accurate results can be obtained using pair approximations \cite{mata2014heterogeneous}. While the nature of the dynamical process, such as its universality class, is not altered by the inclusion of pair correlations, the resulting theory becomes substantially more accurate when compared with simulations, especially in the regime of sparse networks, where $\langle k \rangle \ll N$.

In addition, for high dissonance, a narrow region appears between the frustration and dominance regimes in which the system converges to an equipartitioned state, $\rho_0 = \rho_A = \rho_B = 1/3$. Fig.~\ref{fig:Dynamic_integration}(g) illustrates the temporal evolution within this region for $\lambda = 0.325$. As in the frustration regime, the ``Conflicted'' fraction collapses after a plateau, but the adoption probabilities do not vanish completely. Consequently, the ``Skeptic'' fraction decreases slowly while the single-rumor fractions increase, until the three fractions become equal.

We can explain the emergence of this equipartitioned state analytically by examining the dynamical equations in the late stages of the integration, when the ``Conflicted'' fraction has vanished ($\rho_{A,B} \approx 0$). Since this state arises from a symmetric evolution of both rumors, we can use the reduced system of Eqs.~\eqref{rho_0_reducido}--\eqref{rho_reducido}, obtained in Section~\ref{subsec2} for $\rho_A = \rho_B \equiv \rho$ and $\omega_A = \omega_B \equiv \omega$. Immediately after the collapse of the ``Conflicted'' state, the ``Skeptic'' fraction exceeds the single-rumor fractions ($\rho_0 > \rho$), so that Eq.~\eqref{rho_reducido} gives $d\rho/dt > 0$ for any $\omega > 0$. At the same time, Eq.~\eqref{rho_0_reducido} gives $d\rho_0/dt < 0$ if $\omega > \frac{\alpha \, (1 - \rho_0 - 2\rho)}{2 \, \rho_0 \, (1+\lambda) \, (1-\rho_0-\rho)}$, a condition that is satisfied by any $\omega > 0$ once $\rho_{A,B} = 1 - \rho_0 - 2\rho \approx 0$. Agents therefore keep flowing from the ``Skeptic'' state to the single-rumor states until $\rho_0 = \rho$, that is, until $\rho_0 = \rho_A = \rho_B = 1/3$. Meanwhile, the small flux of agents through the ``Conflicted'' state keeps raising the thresholds (Eq.~\eqref{Mu_dynamic}), so that $\omega$ eventually vanishes. The equipartitioned state is thus a particular case of the frustration fixed point $\vec{\rho}^{\,F}$, reached asymptotically.

In the agent-based simulations, this deterministic band of exact equipartition is replaced by a stochastic crossover along the boundary between single-rumor dominance and frustration. Across this transition, individual realizations may either break symmetry toward one rumor or freeze at a frustrated configuration whose final ``Skeptic'' fraction grows continuously with $\lambda$; consequently, particular boundary realizations can exhibit comparable fractions of skeptics and single-rumor believers (as in Fig.~\ref{fig:Simulation_results}(f)), where $\rho_0$ competes in magnitude with $\rho_A$ and $\rho_B$ before skepticism dominates at larger $\lambda$. However, whereas the homogeneous mean-field integration locks an entire interval of parameters into $\rho_0 = \rho_A = \rho_B = 1/3$ via a slow collective drift sustained by a continuous global threshold distribution, discrete and heterogeneous local neighborhoods in the network prevent this state from forming a separate macroscopic plateau in the ensemble average.

\section{Conclusions}\label{sec5}

In this work, we proposed an agent-based model, built upon the Watts linear threshold framework~\cite{Watts_Model}, to investigate the propagation of correlated rumors driven by social influence. A key feature of our approach is that agents are allowed to adopt opposing rumors simultaneously. Because these narratives are fundamentally opposed, simultaneous adoption induces cognitive dissonance, which agents eventually resolve by abandoning both rumors and increasing their critical mass thresholds. The interplay between the spreading dynamics and this dissonance-driven deactivation yields three macroscopic regimes. In the single-rumor dominance regime, one rumor spreads through the entire network and drives the other to extinction. In the frustration regime, the repeated cycles of adoption and deactivation raise the critical mass thresholds until propagation stops, leaving the ``Skeptic'' state as the most populated one and smaller groups of single-rumor believers. In the dynamic coexistence regime, which emerges for high correlation and low dissonance, the thresholds of practically all agents saturate, but the cumulative influences remain high enough for agents to readopt the rumors, so that they keep cycling between states and a large fraction of the population remains in the ``Conflicted'' state. We showed that single-rumor dominance emerges from a delicate balance between the correlation of the rumors, the rate at which agents leave the ``Conflicted'' state, and the threshold penalty imposed after each deactivation, which shrinks the dominance region as it increases. Dynamic coexistence, in contrast, is insensitive to this penalty and is determined by the correlation and the dissonance alone.

To study the model in greater depth, we derived the mean-field evolution equations for the fractions of agents in each state and analyzed the stability of the fixed points associated with the observed regimes. The state of complete skepticism, $\vec{\rho}^{\,0}$, is unstable, which explains why the system does not converge to the extinction of beliefs. The single-rumor dominance points, $\vec{\rho}^{\,A}$ and $\vec{\rho}^{\,B}$, are stable when the competing rumor can no longer be adopted, and the frustration points, $\vec{\rho}^{\,F}$, are stable once the cumulative increase of the thresholds makes the adoption probabilities vanish ($\omega_A = \omega_B = 0$), freezing the system in a fragmented state. Finally, the dynamic coexistence point, $\vec{\rho}^{\,C}$, exists and is stable when the correlation exceeds a minimum value that increases with the dissonance rate, $\lambda \ge 2\alpha/(2+\alpha)$, with a fraction of ``Conflicted'' agents that depends only on $\alpha$. Since neither this point nor its existence condition depends on $\delta$, this result also explains the insensitivity of the dynamic coexistence regime to the threshold penalty.

The numerical integration of the mean-field equations qualitatively reproduces the three regimes observed in the agent-based simulations, as well as their arrangement in the parameter space, although the extent of the regions differs; in particular, the dynamic coexistence region, bounded by the existence condition of $\vec{\rho}^{\,C}$, extends to higher dissonance values. The integration also exhibits a narrow band at the dominance--frustration boundary that converges deterministically to an equipartitioned frustrated state ($\rho_0 = \rho_A = \rho_B = 1/3$). While individual stochastic realizations near the transition boundary can reach comparable fractions as skepticism begins to compete with single-rumor adoption, the sharp equipartitioned band in the mean-field integration stems from its continuum global threshold distribution, highlighting the role of local neighborhood discreteness and dynamical pair correlations. Future work should therefore incorporate pair approximations and a more detailed representation of the threshold distribution, accounting for the heterogeneous, discrete evolution of individual thresholds rather than relying on a uniform shift of the global distribution.

Finally, this model shows that stable collective regimes of single-rumor dominance, frustration or dynamic coexistence can emerge even in the absence of external changes. Moreover, the formulation is general enough to be extended to other contexts involving mutually related or competing states, such as comorbidities or problematic consumption in health systems, where state adoption or persistence depends on both social influence and the concurrent use of multiple substances.

\section*{Acknowledgments}

FDC and LG were supported by doctoral fellowships from the National Scientific and Technical Research Council (CONICET, Argentina) [Fellowship IDs 192341 and 184217, respectively] and by ``Move la Am\'erica'' exchange fellowships provided by the Coordination for the Improvement of Higher Education Personnel (CAPES, Brazil) [Fellowship IDs 88881.996745/2024-01 and 88881.996200/2024-01, respectively]. SCF acknowledges the financial support by the Funda\c{c}\~ao de Amparo \`a Pesquisa do Estado de Minas Gerais (FAPEMIG)-Brazil [Grant No. APQ-01973-24], Conselho Nacional de Desenvolvimento Cient\'ifico e Tecnol\'ogico (CNPq)-Brazil [Grants No. 407871/2025-0 and 310984/2023-8], and Funda\c{c}\~ao de Amparo \`a Pesquisa do Estado de S\~ao Paulo (FAPESP) [Grant No. 25/24366-1].

\section*{Author contributions}

All authors contributed to the study conception and design. The theoretical analysis was performed by Favio Di Ciocco and Lucio Garcia. The simulations were performed by Favio Di Ciocco. The first draft of the manuscript was written by Favio Di Ciocco and Lucio Garcia, and all authors commented on previous versions of the manuscript. All authors read and approved the final manuscript.

\section*{Competing interests}

The authors have no relevant financial or non-financial interests to disclose.

\section*{Data availability}

The datasets generated and analyzed during the current study are available from the corresponding author on reasonable request.

\bibliographystyle{unsrtnat}
\bibliography{references}

\end{document}